\documentclass[journal=jctcce,manuscript=article]{achemso}
\usepackage{amsmath,psfrag,graphicx,pstricks,caption,url,color}
\usepackage{amsfonts,amssymb,graphicx,wrapfig} 
\usepackage{algorithm,algpseudocode,tikz, verbatim,subcaption}
\usepackage{textcomp} 
\usepackage{gensymb} 
\usepackage[justification=centering]{caption}
\usetikzlibrary{arrows,shapes}

\usepackage{titlesec}
\usepackage{blindtext}
\usepackage[T1]{fontenc}
\usepackage[utf8]{inputenc}
\usepackage{wrapfig}
\usepackage{amsmath}
\usepackage{amssymb}
\usepackage{lastpage}
\usepackage{graphicx}
\usepackage{dcolumn}
\usepackage{bm}
\usepackage{tabularx} 
\usepackage{siunitx}
\usepackage{lineno}
\usepackage{color}
\usepackage{longtable}
\usepackage[normalem]{ulem}
\definecolor{darkgreen}{rgb}{0.0, 0.5, 0.0}

\newcommand{\rc}{r_c}

\newcommand{\kB}{k_{\mathrm{B}}}
\newcommand{\kT}{\kB T}
\newcommand{\myvec}[1]{\vec{#1}}

\newcommand{\rvec}{\myvec{r}}

\title{\textbf{The Role of Chemical Heterogeneity in Surfactant Adsorption at Solid-Liquid Interfaces}}

\author{Jason Klebes}
\affiliation[IBM]{IBM Research Europe, The Hartree Centre, Daresbury, WA4 4AD, United Kingdom}
\alsoaffiliation[Leeds]{School of Mathematics, University of Leeds, LS2 9JT, United Kingdom}
\author{Sophie Finnigan}
\affiliation[IBM]{IBM Research Europe, The Hartree Centre, Daresbury WA4 4AD, United Kingdom}
\alsoaffiliation[Imperial]{Department of Chemistry, Molecular Sciences Research Hub, White City Campus, Imperial College London, Wood Lane, London W12 0BZ, United Kingdom}
\affiliation[STFC]
{The Hartree Centre, STFC Daresbury Laboratory, Warrington, WA4 4AD, United Kingdom}
\author{David J. Bray}
\affiliation[STFC]
{The Hartree Centre, STFC Daresbury Laboratory, Warrington, WA4 4AD, United Kingdom}
\author{Richard L. Anderson}
\affiliation[STFC]
{The Hartree Centre, STFC Daresbury Laboratory, Warrington, WA4 4AD, United Kingdom}
\author{William C. Swope}
\affiliation[IBM]
{IBM Almaden Research Center, San Jose, California, United States}
\author{Michael A. Johnston}
\affiliation[IBM]
{IBM Research Europe, Dublin, Ireland}
\author{Breanndan O Conchuir}
\affiliation[IBM]
{IBM Research Europe, The Hartree Centre, Daresbury WA4 4AD, United Kingdom}
\email{breanndan.conchuir@ibm.co.uk}

\date{}

\begin{document}

\maketitle

\begin{abstract}
Chemical heterogeneity of solid surfaces disrupts the adsorption of surfactants from the bulk liquid. While its presence can hinder the performance of some formulations, bespoke chemical patterning could potentially facilitate controlled adsorption for nanolithography applications. Although some computational studies have investigated the impact of regularly patterned surfaces on surfactant adsorption, in reality many interesting surfaces are expected to be stochastically disordered and this is an area unexplored via simulations. 
In this paper we describe a new algorithm for the generation of randomly disordered chemically heterogeneous surfaces and use it to explore the adsorption behaviour of four model nonionic surfactants. 
Using novel analysis methods we interrogate both the global surface coverage (adsorption isotherm) and behaviour in localised regions. 
We observe trends in adsorption characteristics as surfactant size, head/tail ratio, and surface topology are varied and connect these to underlying physical mechanisms. 
We believe that our methods and approach will prove useful to researchers seeking to tailor surface patterns to calibrate nonionic surfactant adsorption.

\end{abstract}

\section{Introduction}

The adsorption of surfactant molecules onto surfaces is a key factor for a wide range of industrial applications\cite{striolo2017,Rosen1989} such as crop protection \cite{Sonntag1988}, corrosion and fouling resistance\cite{Malik2011}, and food additives\cite{Campos2013}.  
For many applications, the adsorption of surfactants at the solid-liquid interface is a fundamental component of its functionality\cite{paria2004}.  
Surfactants can be utilised to stabilise nanoparticle dispersions\cite{Islam2003,Iijima2009,Lotya2009,Heinz2017,Ma1992a,paria2004,Ma1992b,Palla2000}, remove dirt from textiles\cite{Paria2003}, filter ultrafine particles \cite{Kang1997}, separate mineral ore particulates\cite{Scamehorn1988}, enhance carbon regeneration\cite{Scamehorn1988}, manufacture composite materials \cite{Gong2000,Yamaguchi2004} and deink paper and plastic films\cite{Gecol2001}. The controlled adsorption of surfactants onto flat substrates facilitates admicellar polymerisation\cite{Marquez2007,See2003} and bespoke lithography processes\cite{Wu2004,Jung2004}.
On the other hand, this phenomenon can also be an impediment to the efficiency of some industrial processes\cite{Kamal2016,Iglauer2010}. For example, in Enhanced Oil Recovery (EOR), the role of surfactants is to migrate to the oil-water interface and tune the interfacial tension to maximise the extraction of oil deposits. The adsorption of surfactants onto porous mineral surfaces prevents them from reaching this interface and therefore reduces the overall effectiveness of the process.  
In the design of these products and processes, understanding and predicting surfactant-surface interactions is of the foremost importance.

A wide variety of experimental techniques have been applied to probe the structure of adsorbed surfactants on solid surfaces\cite{Atkin2003, striolo2017, striolo2019}. These include neutron reflectivity\cite{Zhao2009}, neutron scattering\cite{Penfold2012}, ellipsometry\cite{Tiberg1994a,Tiberg1994b,Keddie2001}, nuclear magnetic resonance (NMR)\cite{Schoenhoff2013}, atomic force microscopy (AFM)\cite{striolo2017, hamon2018, Manne1994, Manne1995} and quartz crystal microbalance with dissipation (QCM-D)\cite{Thavorn2014, Shi2009}. A total of six distinct adsorbate morphologies have been observed on flat solid surfaces\cite{striolo2017}. Spheres, cylinders, and flat bilayers were detected on hydrophilic surfaces whereas hemispheres, hemicylinders, and flat monolayers have been observed on hydrophobic surfaces. In another paper by Shi and coworkers\cite{Shi2009}, the authors demonstrated that molecular compounds such as toluene, phenol, and 1-hexanol influence the morphology of adsorbed micelles. 

Despite the extensive collection of experimental studies in the literature, such techniques struggle to extract crucial microscopic information such as the precise structural characteristics and chemical composition of adsorbed micelles\cite{Shi2009}. In contrast, computational techniques allow researchers to not only interrogate the microscopic adsorption of surfactants onto solid surfaces at the atomic resolution, but also investigate the influence of various different surface features and environmental conditions with ease.  Molecular dynamics has been used to simulate the behaviour of surfactants at the solid-liquid interface under simple shear flow.\cite{bradleyshaw2016} Meanwhile, other computational efforts have focused on surfactant adsorption onto assorted surfaces such as alumina\cite{Liu2014}, carbon nanotubes\cite{striolo2017, Tummala2009, Tummala2010} and silica\cite{Tummala2011}.

In recent years, there has been an increasing trend moving towards applying coarse-grained molecular simulation methodologies such as Dissipative Particle Dynamics (DPD) to simulate surfactant adsorption\cite{Taddese2020, Min2012}. The advantages of employing this technique is that it allows one to simulate significantly larger spatial domains and longer time-scales compared to atomistic molecular dynamics\cite{hoogerbrugge_1992, grootwarren, espanol_warren_2017}. In particular, one can observe the restructuring of surfactants within micelles, sample high surfactant concentrations,
and cooperative adsorbed micelle interactions, albeit with some loss of chemical detail\cite{suttipong2016, conchuir2020, striolo2017}.

The majority of these computational studies consider surfaces to be flat, homogeneous structures. In reality, however, both natural and industrial surfaces are almost always rough and chemically heterogeneous. Recent reviews by Striolo et al. have surveyed the profound effect which surface features can have upon surfactant-surface interactions\cite{striolo2019, striolo2017}. They are known to significantly impact surfactant adsorption isotherms\cite{Wu2011,Zhang2007,Cases1992,Leermakers2005,hamon2018,Lajtar1993,Zhou2019}, the heat of adsorption\cite{Lajtar1993,Narkiewiczmichalek1993}, the number of adsorbed surfactants\cite{Wu2011,Foisner2004,Reimer2005,Lajtar1993,suttipong2014} and adsorbed micelle morphologies\cite{Sammalkorpi2008,Foisner2004,Reimer2005,Zhang2007,Tummala2010,Leermakers2005,hamon2018,suttipong2014}. 

From a modelling and simulation perspective, several authors have attempted to directly link the topology of chemical surface heterogeneity to the adsorption characteristics of surfactant dispersions. Suttipong and coworkers have been particularly prominent, publishing a series of DPD papers on a variety of different surface patterns\cite{suttipong2014,suttipong2015,suttipong2017}. In the first paper, they interrogated the effects of lateral confinement and cooperative adsorption by simulating surfactant assembly on single and parallel hydrophobic stripes surrounded by repellent hydrophilic regions\cite{suttipong2014}. Next, they examined the intricate interplay between geometrical and chemical properties by simulating surfactant assembly in surface trenches\cite{suttipong2015}. Finally, the adsorption of surfactants on perpendicular intersecting hydrophobic stripes and hydrophobic steps was studied\cite{suttipong2017}.

Meanwhile, Reimer et al.\cite{Reimer2005} and Zhang et al.\cite{Zhang2007} employed the lattice Monte Carlo technique to simulate adsorption on checkerboard and striped hydrophilic-hydrophobic surfaces. The former discovered that the size of the adsorbed structures was determined by the dimensions of the surface features, while the latter found that the measured absorption isotherms can be used to infer information about the heterogeneous surface patterns.

Although these results provide invaluable insights into the complex relationship between surfactant adsorption and chemical heterogeneity, a full understanding of how different surfactants interact with increasingly realistic model surfaces of varying chemical composition remains an important field of study.
In this article, we design the first mathematical algorithm for the generation of randomly disordered chemically heterogeneous surfaces. 
DPD simulations are then conducted for varying concentrations of four model nonionic surfactants in aqueous solutions. Novel analysis techniques are derived and applied to directly connect the surface topologies to the measured global surface coverage and the localised adsorption density distributions.

The remainder of this article is arranged as follows. First, we provide an overview of the surface generation algorithm along with the specific surfaces and molecules simulated in this study. Next, we detail our DPD simulation protocol and applied adsorption metrics. We then discuss the results of our computations, where we examine the role of surfactant polarity and localised adsorption features. The paper concludes with a discussion on how this model can be exploited to tailor surfactant adsorption at solid-liquid interfaces. 

\section{Developing the DPD Model}

In this work we adopt the Dissipative Particle Dynamics (DPD) approach in which  particles (DPD beads) interact via soft repulsions and with local pairwise dissipative and random forces which together work as a thermostat.\cite{grootwarren} Rather than repeat here the details of what is now a fairly standard simulation method, we point the reader to the Supporting Information (SI), chapter 17 of the textbook by Frenkel and Smit,\cite{frenkelsmit_2002} and the original DPD literature.\cite{hoogerbrugge_1992,grootwarren,espanol_warren_1995} For an up to date perspective on the DPD methodology see Espa\~nol and Warren. \cite{espanol_warren_2017} In our DPD simulations we use a standard choice for reduced units in which the beads have unit mass, the system is governed by temperature $\kT=1$, and the baseline cut-off distance for the short-range soft pairwise repulsions between solvent beads is set as $\rc=1$.

\subsection{Coarse Grained Molecule Development}

\begin{figure}[ht]
    \centering
    \includegraphics[width=\textwidth]{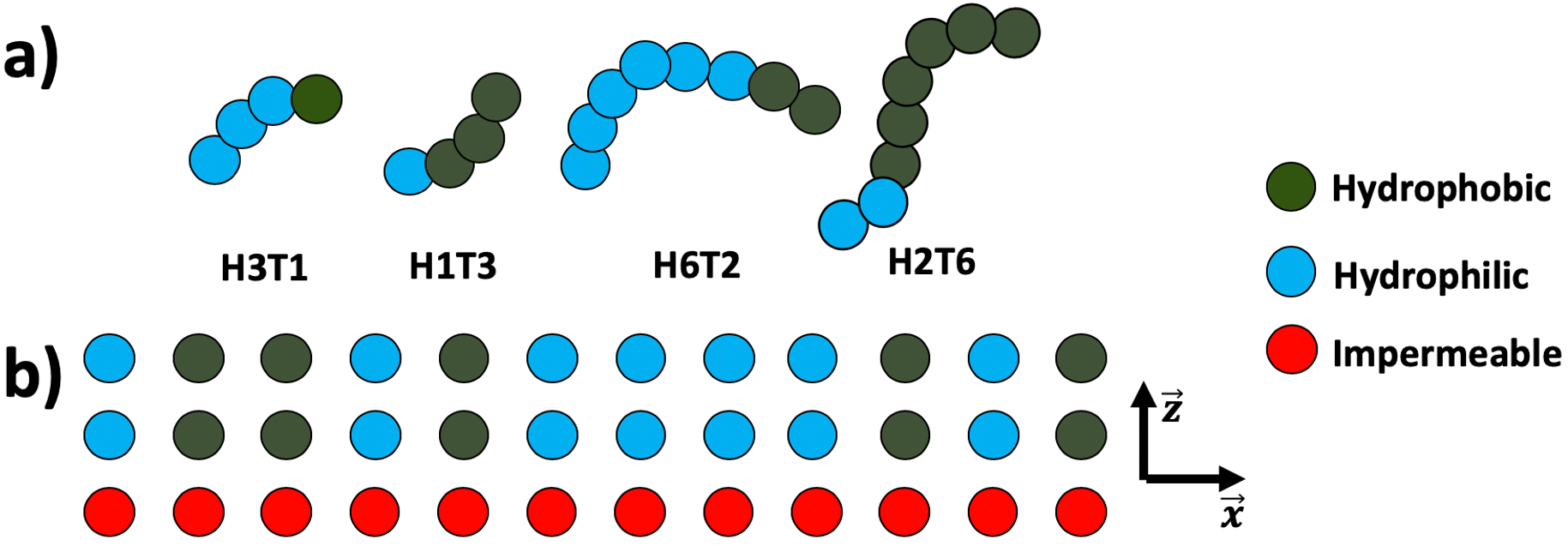}
    \caption{Schematic of a) the four surfactant molecules H3T1, H1T3, H6T2, and H2T6 and b) surfaces simulated in this study.}
    \label{fig:molecules}
\end{figure}

Four model nonionic surfactant molecules were explored in this investigation, as illustrated in Fig. \ref{fig:molecules}a). These linear molecules consist of hydrophobic alkane tail (T) beads and hydrophilic polyethylene oxide head (H) beads in varying proportions. 
We chose to simulate the adsorption behavior of two different H-T ratios - 1:3 and 3:1 - and two different lengths - 4 beads and 8 beads.
Henceforth, our model surfactants will be referred to as $\mathrm{H1T3}$, $\mathrm{H3T1}$, $\mathrm{H2T6}$ and $\mathrm{H6T2}$ (Fig. \ref{fig:molecules}a). Note that these specific molecules were chosen because their H-T ratios and lengths allow us to probe a broad variety of potential surfactant architectures.
In our model, water beads are representative of (on average) two water molecules. The corresponding volume of these beads is approximately $30$ \AA$^{3}$. Adopting the now somewhat standard reduced density for DPD $\rho = 3$, \cite{grootwarren} this sets the coarse grained length scale as 1 reduced DPD length unit, r$_c$ = 5.65 \AA. 

Surfactant molecules are connected via simple harmonic bonds and angles between constituent beads. The expression for the harmonic radial bond $U(r_{i,j})$ between each neighbouring pair of molecular beads $i$ and $j$ is written in Eq. \ref{eq:bonds}\cite{dlmeso}. Here the spring constant $\kappa_{r}$ was set to 150 and the cutoff distance $r_{0}$ set to $0.44/0.56/0.68$ for T-T/T-H/H-H bonds, respectively, as dictated by the force field parameterisation protocol detailed in the Supporting Information. 
Similarly, there exists a harmonic angular bond $U(\theta_{ijk})$ for each set of three consecutive beads along the molecule chain. The equilibrium bond angle $\theta_{0}$ is set to 180$\degree$ for all angular bonds, along with a stiffness parameter $\kappa_{\theta}$ set to 5 following the pragmatic approach of Anderson et al. \cite{anderson_2017,anderson_2018}
\begin{equation}
\begin{aligned}
U(r_{ij}) &= \frac{\kappa_{r}}{2}(r_{ij} - r_{0})^{2}, \quad U(\theta_{ijk}) &= \frac{\kappa_{\theta}}{2}(\theta_{ijk} - \theta_{0})^{2}.
\end{aligned}
\label{eq:bonds}    
\end{equation}

\subsection{Coarse Grained Surface Generation} 
\label{ssec:surfacegeneration}

Our template surface consists of a three layer deep cuboidal lattice of beads with a pre-specified inter-bead separation, $a$ (Fig. \ref{fig:molecules}b) 
The bottom layer consists of strongly repulsive `impermeable' beads (IMP) whose function is to prevent the unphysical phenomenon of molecules penetrating into the surface. The middle layer is generated as outlined further in this section
with identical hydrophilic (T) and hydrophobic (H) beads to those employed in the surfactant molecules. Finally, the configuration of the middle layer is replicated in the top, exposed, layer. Three layers was found to be the thinnest DPD solid surface template thickness which remains impermeable to solvent and surfactant molecules. We note that the permeability of the surface is a consequence of the soft potentials adopted in DPD. 

\begin{figure}[ht]
    \centering
    \includegraphics[width=\textwidth]{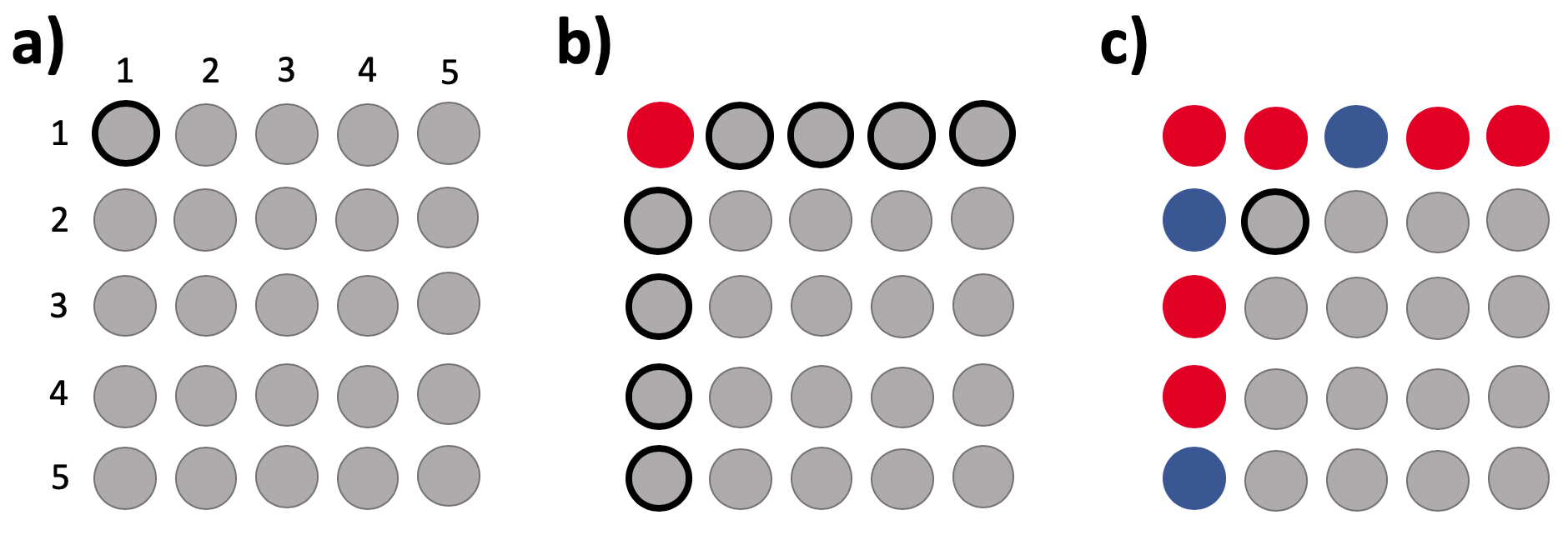}
    \caption{Schematic of the generation of disordered surfaces of hydrophobic (blue) and hydrophilic (red) DPD beads. The bead type is first assigned at a) one of the corners $b[1,1]$, followed by b) the first row and column and finally c) for the remaining beads.}
    \label{fig:surfGen}
\end{figure}

An algorithm was developed to generate bespoke hydrophobic-hydrophilic surface patterns, as a function of three geometric parameters.
The disordered surfaces simulated in this paper are constructed by defining two independent generating probabilities $p_{x}$ and $p_{y}$ where $0\leq p_{x},p_{y} \leq 1$ and target surface area fraction $s_{tar}$. The allocation of surface beads is implemented using a stochastic Monte Carlo approach as illustrated in Fig.~\ref{fig:surfGen}. 

The first corner bead type ($b[1,1]$ in Fig.~\ref{fig:surfGen}a) is defined by randomly sampling from the global target hydrophobic-hydrophilic surface area fraction $s_{tar}$. 
Next, the first row and column (Fig. ~\ref{fig:surfGen}b) are sequentially generated based upon the generating probabilities $p_x$, $p_y$ and the preceding bead type allocation. Finally, the identities of the remaining surface beads are systematically assigned, according to the following algorithm:

\begin{equation}
b[i,j] =
\begin{cases}
b[i-1,j], & r_{x}<p_{x},r_{y}\geq p_{y} \\
b[i,j-1], & r_{x}\geq p_{x},r_{y}<p_{y} \\
b[i-1,j], & r_{x}<p_{x},r_{y}<p_{y},p_{x} - r_{x} \geq p_{y} - r_{y}\ \\
b[i,j-1], & r_{x}<p_{x},r_{y}<p_{y},p_{x} - r_{x} < p_{y} - r_{y}\ \\
r_{tar}, & r_{x}\geq p_{x}, r_{y}\geq p_{y}, 
\end{cases}
\label{eq:algorithm}
\end{equation}
where $r_{tar}$ is a random sample from $s_{tar}$.

After a surface is generated, an extra check is performed to ensure that the stochastically generated hydrophobic-hydrophilic surface area fraction is within a tolerance cutoff $\delta_{c}$ of the target value $s_{tar}$. Surfaces are recursively formulated until this tolerance criterion has been satisfied

The algorithm outlined in Eq. \ref{eq:algorithm} also defines a two dimensional patch length distribution $f(l_{x},l_{y})$.  Here, $l_{i}$ signifies the length of sequences of the same bead type in axial direction $i$. This function can be decomposed into three components:
\begin{equation}
f(l_{x},l_{y}) = f_{x}(l_{x}) + f_{y}(l_{y}) + f_{\sigma}(l_{x},l_{y}).
\end{equation}
Here $f_{x}(l_{x})$ and $f_{y}(l_{y})$ represent the mutually independent length distributions in the $x$ and $y$ axes. The term $f_{\sigma}(l_{x},l_{y})$ denotes the effects of axial correlation on the overall distribution. Note, that the length probability distributions $f_{x}(l_{x})$ and $f_{y}(l_{y})$ are geometric distributions. However, mapping the surface from a discrete computational lattice onto a continuous spatial domain transforms these functions into exponential distributions $F(l_i)$. 
By the central limit theorem, the global average of the stochastic localised correlation effects is negligible for large surfaces. As a result, in the continuum limit $F_{x}(l_{x})$ and $F_{y}(l_{y})$ can be considered to be independent distributions.

The exponential distribution $F$ has the form:
\begin{equation}
 F(l_{i}) \propto \exp\bigg(\frac{-l_{i}}{d^b_{i}}\bigg) , \quad d_i^b=\dfrac{-1}{\ln(P^b_{i})} , \quad i\in\{x,y\},
\label{eq:expdis}
\end{equation}

While the generating probability $p_i$ describes a choice between continuing with the same bead type or a randomly chosen bead, we here use the probability $P_i^b$ of choosing the same bead type, $b$, rather than the opposite bead type at each step.  This probability follows from the generating probability as:  
\begin{equation}
P^b_{i}=p_i+s_{tar}^b(1-p_i),
\label{eq:probability}
\end{equation}
where $s^{b}_{tar}$ signifies the specific target area fraction for bead type $b$.
 
Starting from any bead, the probability distribution of continuing with the same type for $N$ beads in direction $i$ is proportional to the exponential distribution outlined in Eq. \ref{eq:expdis}.
The expectation value $\langle N \rangle$, i.e. the mean number of beads a domain is expected to continue with the same bead type for, before encountering a bead of the opposite type, is  
 \begin{equation}
    \langle N_i^b \rangle =\dfrac{\int l_i e^{-l_i/d_i^b} dl_i }{\int e^{-l_i/d_i^b} dl_i} 
\label{eq:meanlength}
\end{equation}
Integrating, the result is identical to the decay length $\langle N_i^b \rangle = d_i^b$.  Therefore the length $d_i^b=\dfrac{-1}{\ln(P^b_{i})}$ is identified with the mean domain size for bead type $b$.

\subsection{Simulated Surfaces}
Three distinct classes of surfaces were generated and examined in this study. Each one has spatial dimensions of $40\times40$ DPD units and consists of $63\times64$ beads. This corresponds to a lattice spacing $a_{x,z}=0.625$ and $a_y=0.635$.

The first set consists of two surfaces, a purely hydrophobic and a purely hydrophilic surface. The simulated adsorption behaviour of surfactants onto these surfaces acts as a reference point for the behaviour on patterned surfaces.

The second set of surfaces contain randomly generated heterogeneous patterns with hydrophilic and hydrophobic patches of various sizes. Each surface was built with a target hydrophilic : hydrophobic surface area fractions $s^{T}_{tar}:s^{H}_{tar}$ of $0.5:0.5$.  

\begin{figure}[ht]
    \includegraphics[width=\textwidth]{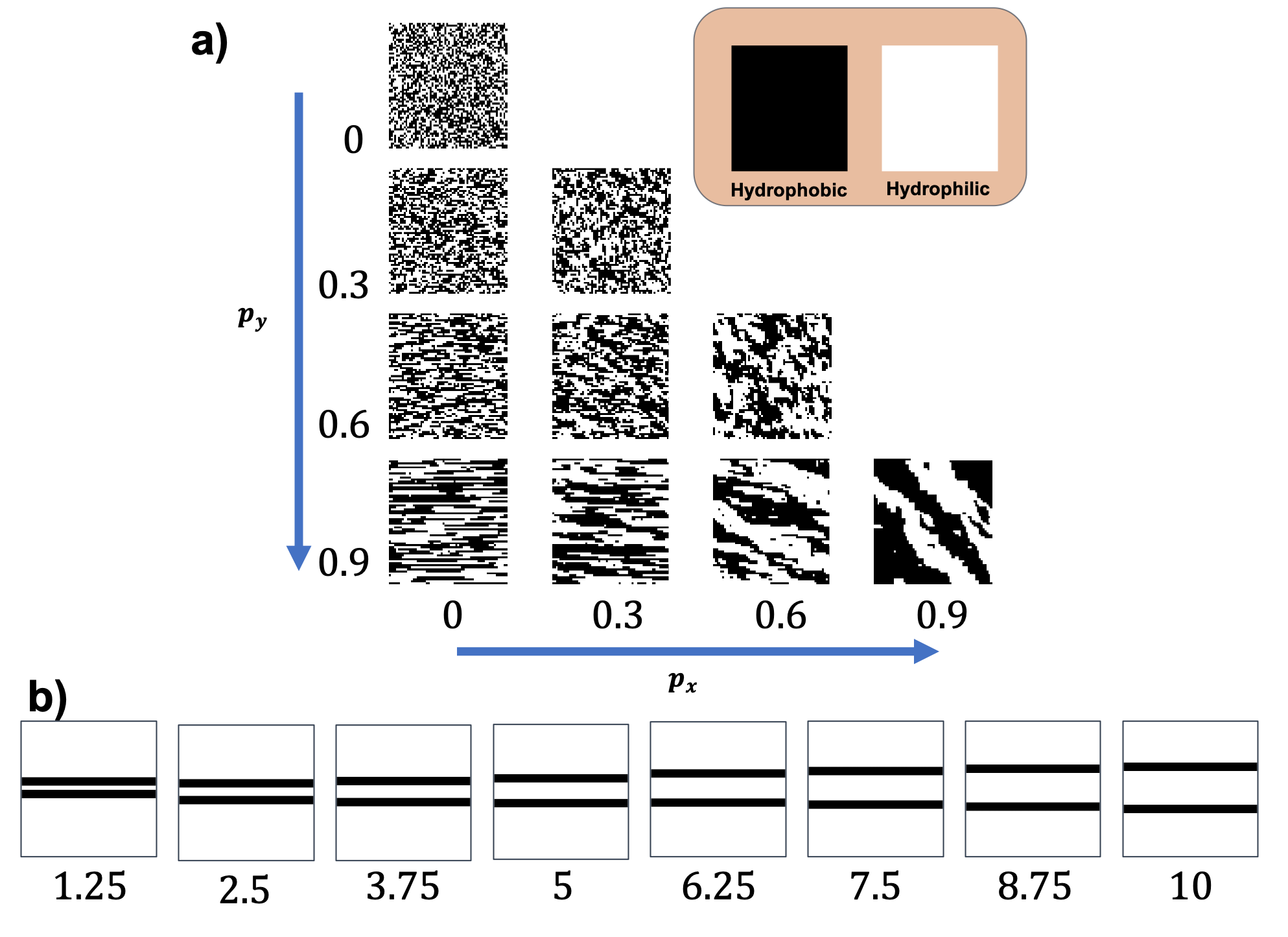}
    \caption{Schematic of a) the ten randomly generated disordered surfaces simulated in this study and b) the eight surfaces with two horizontal hydrophobic stripes. The purely hydrophilic and purely hydrophobic surfaces are also illustrated. Each square pixel in the images represents one surface bead.}
    \label{fig:surfaces}
\end{figure}

A discrete set of four generating probabilities $p_x, p_y\in \{0, 0.3, 0.6 ,0.9\}$ was chosen.  The symmetry of this parameter space allows us to reduce it from sixteen to ten distinct possible combinations (Fig. \ref{fig:surfaces}a).  
Note, that the cutoff tolerance $\delta_{c}$ around the target area fraction $s^{H}_{tar}=0.5$ was fixed as $\pm 0.1$.  As such, the actual hydrophobic area fractions range from $45\%$ for surface (0.6, 0.9) to $56\%$ for surface (0.9,0.9).  
Each of these ten surfaces, along with the fully hydrophilic and fully hydrophobic surfaces, was generated once and combined with four different surfactant molecules at the four different solute concentrations.  

The third set of simulations involved surfaces with two horizontal hydrophobic stripes superimposed on a hydrophilic background (Fig. \ref{fig:surfaces}b). These eight surfaces contained stripes of identical width (four surface beads or 2.5 DPD units), while the gap width ranged from two to sixteen beads (1.25 to 10 DPD units). 

The patch length distribution encoded in the generated surfaces defines the spatial heterogeneity of the surface potential encountered by nearby surfactant molecules. The theoretical mean of this distribution (see Eq. \ref{eq:meanlength}) is outlined in Table \ref{tab:TheoryMeanLength} for each of the simulated generating probabilities.

\begin{table}[h!]
  \begin{center}
    \begin{tabular}{c|c}
      Generating probability $p$ & Mean Patch Length  $d$ \\
      \hline
      0 &  0.90 \\
     0.3 &  1.45\\
      0.6 &  2.80\\
      0.9 & 12.19\\
      \end{tabular}
  \end{center}
  \caption{Table of the characteristic patch length scales measured in DPD units.}
  \label{tab:TheoryMeanLength}
\end{table}

\begin{table}[h!]
  \begin{center}
    \begin{tabular}{c|c|c}
      Surface ($p_x, p_y$) & $\langle d_x \rangle$ &  $\langle d_y \rangle$ \\
      \hline
      (0, 0)  & \bf{1.25}&1.26\\
      (0, 0.3) & \bf{1.24} & 1.85\\
      (0, 0.6) & \bf{1.25} & 3.36\\
      (0, 0.9) & \bf{1.23} & 12.41 \\
      (0.3, 0.3) &\bf{1.73} & 1.76\\
      (0.3, 0.6)  &\bf{1.61} &2.97\\
      (0.3, 0.9) & \bf{1.78} & 11.07\\
      (0.6, 0.6)  & 3.00 & \bf{2.83}\\
      (0.6, 0.9) & \bf{3.22} & 8.83\\
      (0.9, 0.9) & \bf{7.61}&8.36 \\
      \end{tabular}
  \end{center}
  \caption{Numerically extracted mean patch lengths of the generated disordered surfaces in DPD units. The shorter axial patch length is highlighted in bold.}
  \label{tab:numMeanLength}
\end{table}

Finite-sized effects account for the variation between the theoretical and numerical mean patch lengths as illustrated in Table 2. The fixed spatial extent of the simulated surfaces results in an amplification of this divergence for more ordered surfaces, where discontinuities across the periodic boundaries, localised correlations between x- and y-direction bead generation and stochastic variation are more significant.

\subsection{Non-Bonded DPD Interactions}

The non-bonded interactions adopted in this work are the usual DPD pairwise soft repulsions, $U_{ij}^C=\frac{1}{2}A_{ij}R_{ij}(1-r_{ij}/R_{ij})^2$ for $r_{ij}\le R_{ij}$ and $U_{ij}^C=0$ for $r_{ij}>R_{ij}$. Here $A_{ij}$ is the repulsion amplitude of the conservative force, $R_{ij}$ the cut-off distance ($R_{ij}$ = $r_c$ = 1), and $r_{ij}=|\rvec_j-\rvec_i|$ the separation between beads $i$ and $j$ located at $\rvec_i$ and $\rvec_j$ respectively. \cite{frenkelsmit_2002, hoogerbrugge_1992,grootwarren,espanol_warren_1995,espanol_warren_2017} The interaction strength between each pair of DPD beads, $A_{ij}$ , is detailed in Table \ref{tab:parameters}. The H and T interaction parameters were parameterized to reproduce the critical micelle concentration (CMC) and mean aggregation number of poly(ethylene oxide) alkyl ethers. Details on how the parameterisation process was performed is provided in the Supporting Information.  

\begin{table}[h!]
\begin{tabular}{|c|c|c|c|c|}
\cline{1-5}
  &  H2O & T & H & IMP\\ \cline{1-5}
 H2O &  25.0 & 51.55 & 29.49 & 70.0 \\ \cline{1-5}
 T &   &37.33  & 40.45 & 70.0 \\ \cline{1-5}
H &  & & 34.19 & 70.0\\ \cline{1-5}
IMP &  &  & & 70.0\\ \cline{1-5}
\end{tabular}
\caption{Table of the set of amplitude prefactor parameters $A_{ij}$ employed in this study.}
\label{tab:parameters}
\end{table}

Identical force field parameters were assigned to both the solute and surface hydrophobic and hydrophilic beads in this study. This allowed us to probe the formulation space where cohesive micelle interactions are of the same order of magnitude as surface interactions, which a priori should produce the broadest range of distinct physical phenomenon as no single interaction dominates. 
 
Note, that this condition should be relaxed in order to generate quantitative predictions for any particular experimental scenario. The interactions of the surface materials would need to be systematically parameterised, similar to the protocol we applied to the bulk system (see Supporting Information for further details).
This exercise is beyond the scope of this work, which is limited to establishing and demonstrating the DPD model itself.

\section{Simulation Protocol}

Dissipative Particle Dynamics (DPD) simulations were carried out in the NVT ensemble using the DL\_MESO simulation package\cite{dlmeso}. In our model, a standard choice for reduced units is adopted in which the beads have unit mass, the system is governed by temperature $\kT=1$ (equivalent to 298 K), and the baseline cut-off distance for the short-range soft pairwise repulsion between solvent beads is set as $\rc=1$. Specific details on the DPD approach and the implementation of surface interactions can be found in the literature\cite{conchuir2020,johnston_2016,dlmeso,espanol_warren_2017}.

We performed a series of simulations interrogating the adsorption behaviour of an aqueous solution of surfactants in contact with a chemically heterogeneous surface. Following well-established protocol\cite{grootwarren}, the density of water beads was set to three beads per unit volume. Each combination of surface and surfactant species was simulated for four distinct molecular concentrations ($0.5\%$, $4.5\%$, $8.5\%$ and $12.5\%$ bead number fraction).

\subsection{Dimensions and Boundary Conditions}
Each simulation box had a length ($x$) and width ($y$) of $40$ DPD units, along with periodic boundary wall conditions along these axes.  In the vertical z-direction, the boundary conditions are hard boundary conditions, as implemented in the DLMESO 2.6 software package \cite{dlmeso}. This boundary condition consists of specular reflection (a reversal of the perpendicular velocity component) along with a repulsive DPD-style quadratic potential with a distance cutoff $z_c = r_c =1$ and wall amplitude prefactor parameter $A_{wall\ i}$. The parameters $A_{wall\ i}$ of the interaction of each species with the boundary were chosen to be the same as those of each species with a water bead, $A_{wall\ i}=A_{\mathrm{H20}\ i}$.  This is intended to reproduce the effect of an extended aqueous bulk phase at the top $z$-direction boundary.  The solid-liquid interface is modelled as a wall of frozen DPD beads at the $z=0$ boundary of the simulation box.  At the bottom $z$-direction boundary, the same hard boundary conditions are only theoretical and have no practical effect, as our surface slab is inserted between the aqueous bulk and this boundary. The simulation boxes extended for a height of $80$ DPD units from the solid-liquid interface to the top $z$-direction boundary. The initial positions of the surfactant and water molecules in the box above the surface were randomly generated.

\subsection{Equilibration}
\label{ssec:equilibration}
A well-established sliding window weighted linear-least squares (WLLS) procedure was implemented to compute the estimated equilibration time (i.e. time at which a steady state value is achieved) of each measured time series observable \cite{johnston_2016}. The global equilibration point was defined as the equilibration point of the slowest converging observable time series. Next, the auto-correlation time was calculated for each time series observable from this point until the end of the simulation. This range of simulation frames is equivalent to the set of equilibrated frames.

Two additional statistical tests were performed to test the stationary property of the equilibrated time series data. If any of the equilibrated time series failed either the KPSS\cite{Kwiatkowski1992} or the Augmented Dickey-Fuller\cite{Dickey1979} stationary tests, then the global equilibration point was incremented forward in time. The new set of equilibrated frames would again be tested for variance using these two tests, and this iterative procedure continued until a global equilibration point was reached where both tests were successful. The mean and standard error 
were recorded for the final set of equilibrated 
simulation frames. 

Each simulation was constructed such that 4 million time steps would complete, which we found was adequate to ensure that the final number of equilibrated frames exceeded at least 5 of the longest auto-correlation timescales, computed from the initial global equilibration point. Similar conditions are commonly imposed such that reliable statistics are recorded from molecular simulations \cite{chodera}. 

\subsection{Adsorption Metrics}

The solid-liquid adsorption behaviour of the studied surfactants was explored using both global and localised metrics.
A molecule was considered adsorbed if any of its constitutive beads was located within $1.3125$ DPD units of the top layer of surface beads.
This heuristic cutoff distance corresponds to half of the surface lattice spacing ($a=0.625$) plus the DPD interaction cutoff distance ($r_c=1$).
While the criterion for the adsorption of a molecule is arbitrarily constructed, a variety of alternative schemes and cutoff distances were observed to yield qualitatively similar results.  
The number of adsorbed molecules was temporally averaged over all equilibrated frames in a simulation, yielding a global adsorption metric for each molecule type and surface pattern at a given concentration.

For the studied homogeneous surfaces, adsorption isotherms were mapped by measuring this quantity across a range of concentrations.
For adsorption isotherms at heterogeneous surfaces, we can construct a baseline theoretical expectation for the total adsorption isotherms $\Theta_t(c)$ by linearly combining the number of molecules adsorbed on separate fully hydrophilic and fully hydrophobic surfaces, weighted by their respective surface area fractions. 
That is to say:
\begin{equation}
    \Theta_t(c) = s^{T}\Theta^{T}(c) + s^{H}\Theta^{H}(c), \quad s^{T} + s^{H} = 1.
\label{eq:theoreticalValue}
\end{equation}
Here $\Theta^{b}(c)$ represents the number of adsorbed molecules on fully hydrophobic/hydrophilic surfaces for a bulk surfactant concentration $c$. The factor $s^{b}$ is the surface area fraction composed of hydrophobic/hydrophilic patches. This equation is a particularly simple discrete and two-component case of the integral adsorption equation, which describes adsorption onto heterogeneous surfaces with varying surface potential. \cite{sips1948,borowko2002, koopal2020}.

In addition to the total number of adsorbed molecules, spatially localised adsorption densities were also measured. The local adsorbed surfactant density was defined as the density of surfactant beads in a rectangular box of dimensions $0.625 \times 0.635 \times 1 $ placed upon the top surface boundary and centred on the position of each individual surface bead. This quantity was temporally averaged over the equilibrated portion of each simulation. 

\section{Results and Discussion}

The following section is divided into two parts. In the first subsection, we analyse the global adsorption properties of the four studied molecules on the generated surfaces. The second subsection is dedicated to the examination of localised adsorption density features and the influence of cooperative and bridging effects.

\subsection{Adsorption Characteristics of Studied Molecules}\label{ssec:hydrophobic}

We begin our investigation by studying the predominantly hydrophilic short-tailed surfactants, H3T1 and H6T2. These molecules do not form micelles in the bulk at the studied concentrations, but diffuse rapidly as individual solvated molecules. They adsorb onto the surface in a mutually independent fashion with the relevant simulated time series typically reaching equilibration
early in the simulation.

\begin{figure}[ht]
    \centering
    \begin{subfigure}[b]{0.43\textwidth}
    \includegraphics[width=\textwidth]{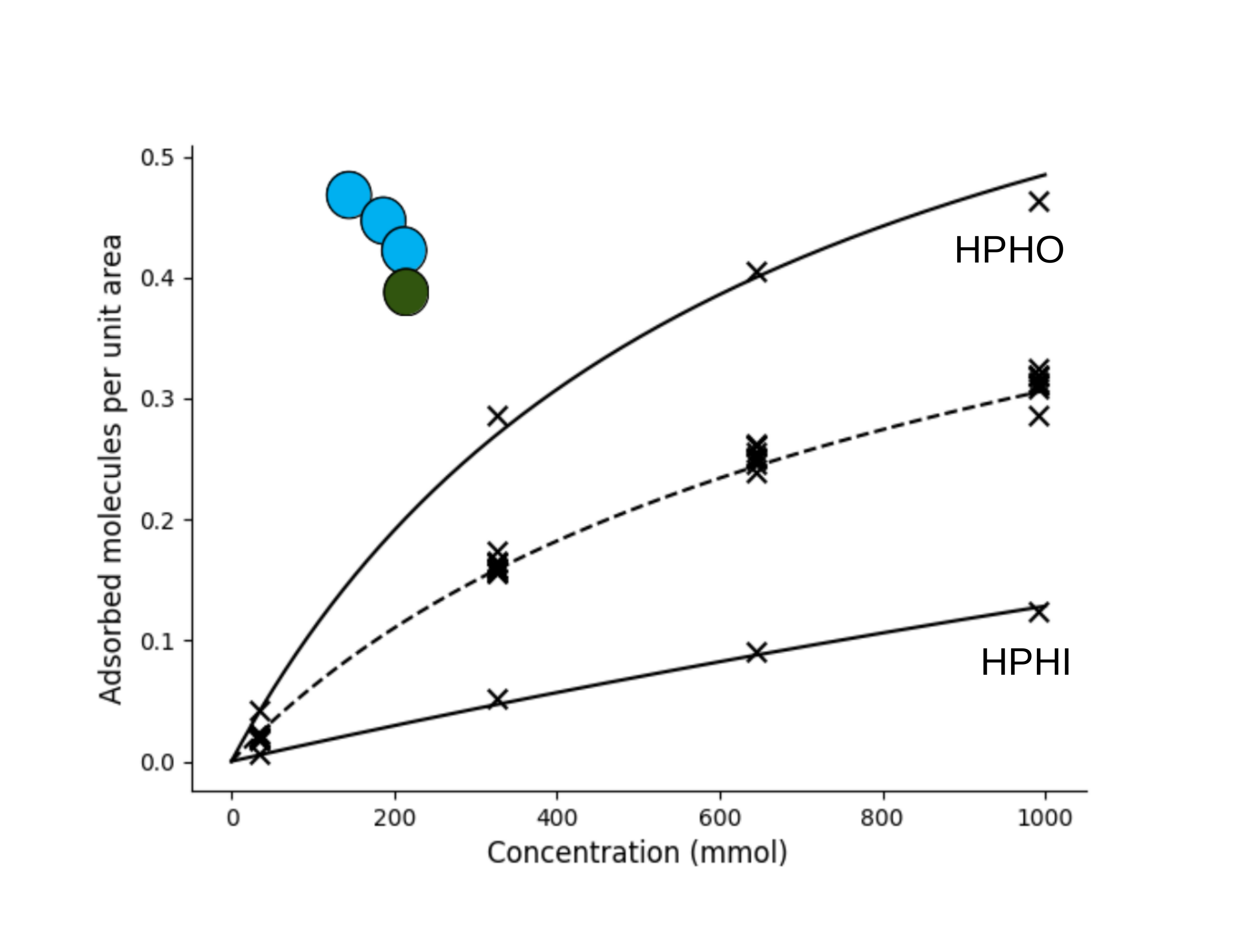}
    \caption{H3T1}
    \end{subfigure}
    ~     \begin{subfigure}[b]{0.43\textwidth}
    \includegraphics[width=\textwidth]{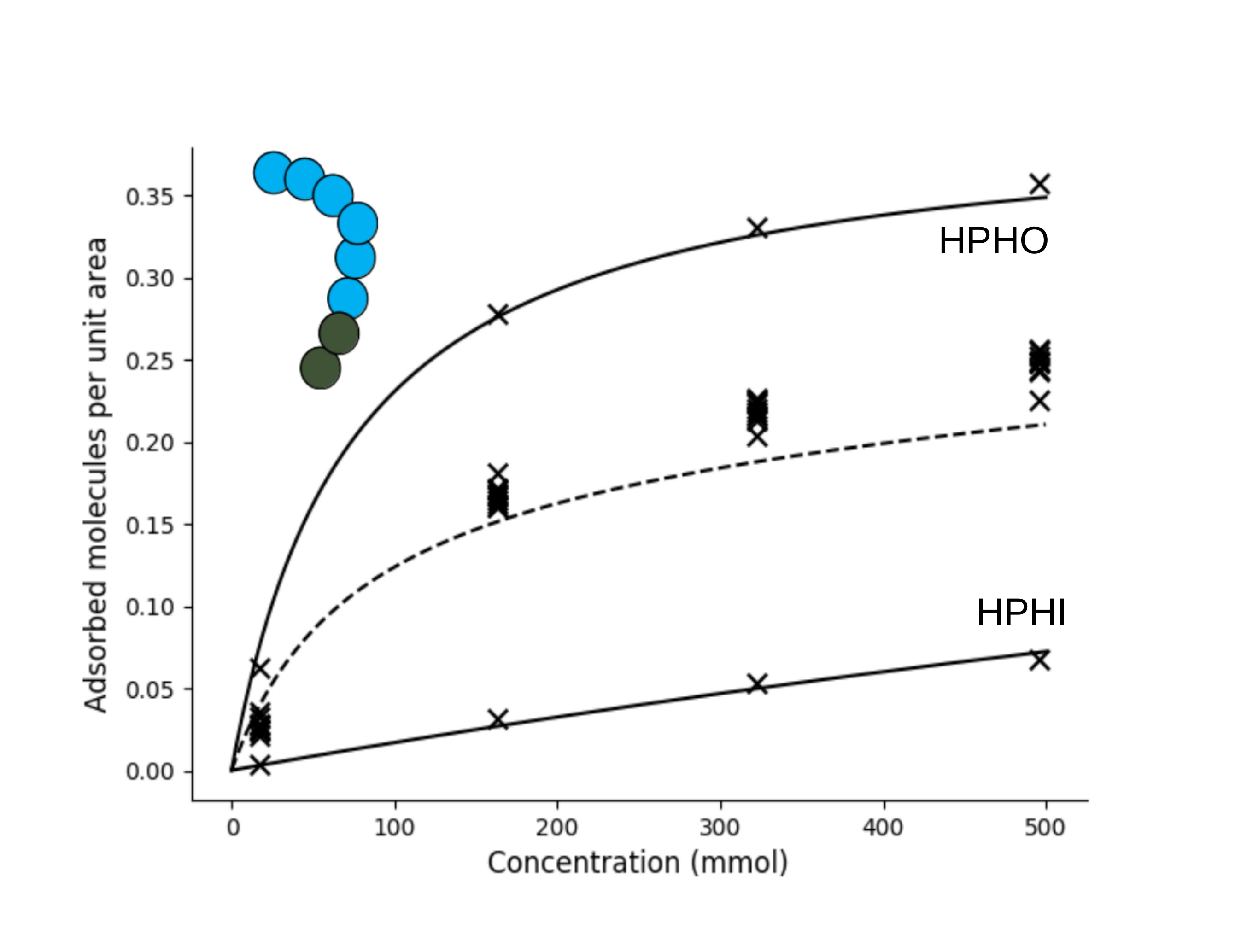}
    \caption{H6T2}
    \end{subfigure}
        \caption{Plots of the adsorption isotherms of molecules H3T1 and H6T2.  The solid curves represent the Langmuir isotherm fitted to the calculated adsorbate density on homogeneous hydrophobic (HPHO) and hydrophilic (HPHI) surfaces, while the dashed curve is the prediction for adsorption at 0.5:0.5 heterogeneous surfaces (See Eq. \ref{eq:theoreticalValue}). The surrounding data points show adsorption in simulations at the ten heterogenous surfaces, calibrated to correct for minor deviations of surfaces area from exactly 50\% hydrophobic area fraction.}
        \label{fig:hphinumads}
\end{figure}

On purely hydrophobic or purely hydrophilic surfaces, we observe that the number of adsorbed molecules rises with increasing concentration. 
This increase follows a Langmuir-type adsorption isotherm \begin{equation}
    \theta(c) = \Gamma_{\infty} \dfrac{1+K c}{K c},
\end{equation} where we fitted the equilibrium parameter $K$ to the number of adsorbed molecules per unit area $\theta(c)$, across four surfactant concentrations.  The parameter $\Gamma_{\infty}$, the high-concentration limit of adsorbate density, was estimated to be $0.79$ molecules per (DPD unit length)$^2$ for H3T1 and $0.40$ for H6T2 from the maximum adsorption observed at high concentrations onto hydrophobic surfaces.  The Langmuir-type isotherm is a good fit for both of the hydrophilic-dominated molecules (Fig. \ref{fig:hphinumads}).

The Langmuir isotherm was originally formulated based on a simplified statistical model of adsorption of point particles onto an array of equivalent surface sites and should not in general be expected to describe adsorption of spatially extended, interacting surfactant molecules from solution \cite{zhu1991}.  Despite the fact that more detailed extensions and alternative theories have been formulated to explicitly model the adsorption of surfactants on solid surfaces \cite{borowko2002}\cite{koopal2020}, we nevertheless find that the basic Langmuir isotherm model describes our data surprisingly well.

This observation is indicative of negligible collective or cooperative effects in the adsorption process. Experimental studies of nonionic surfactant adsorption onto homogeneous surfaces have uncovered a multi-step model, where, for surfactant types or concentrations where cooperative surface structures do not form, adsorption follows a Langmuir isotherm.\cite{zhu1991}\cite{paria2004} Fig. \ref{fig:hphinumads} illustrates that for the surfactants H1T3 and H2T6 within the range of concentrations studied here, we are indeed observing the adsorption of individual molecules and the Langmuir-type isotherm that is characteristic for this regime. 

Adsorption at the heterogenous surfaces is also shown in Figure \ref{fig:hphinumads}.
As outlined in Eq. \ref{eq:theoreticalValue}, a baseline prediction for adsorption onto mixed surfaces with 0.5:0.5 area fraction can be constructed by taking the mean of observed isotherms at the two pure surface types.  This linearly constructed baseline prediction is shown as the dashed line in Fig. \ref{fig:hphinumads}. The associated data points indicate time-averaged adsorption density observed in simulations at each of the ten heterogeneous surfaces. The observed densities were standardized to allow comparison with exactly 50\% hydrophobic surfaces, by dividing by the hydrophobic surface area fraction of each surface, which was in the range 0.45-0.56, and multiplying by $0.5$.  Comparing the data points in Fig. \ref{fig:hphinumads} to the projected adsorption isotherm, we see that the prediction is qualitatively accurate in both cases.  We also observe that the adsorption of the longer molecule H6T2 consistently exceeds the theoretical prediction.  

\begin{figure}[ht]
    \centering
    \begin{subfigure}[b]{0.23\textwidth}
    \includegraphics[width=\textwidth]{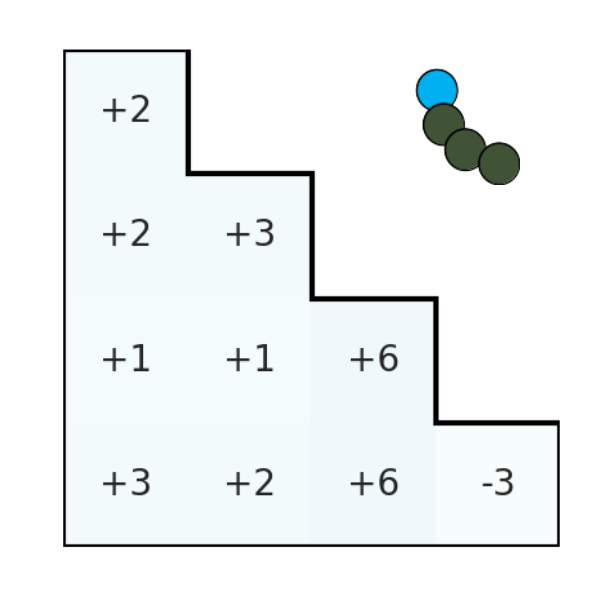}
    \caption{H3T1}
    \end{subfigure}
    ~ 
    \begin{subfigure}[b]{0.23\textwidth}
    \includegraphics[width=\textwidth]{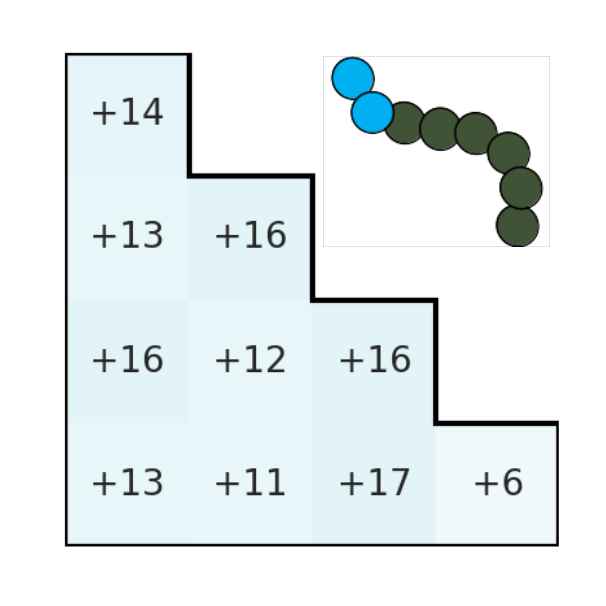}
    \caption{H6T2}
    \end{subfigure}
    \begin{subfigure}[b]{0.23\textwidth}
    \includegraphics[width=\textwidth]{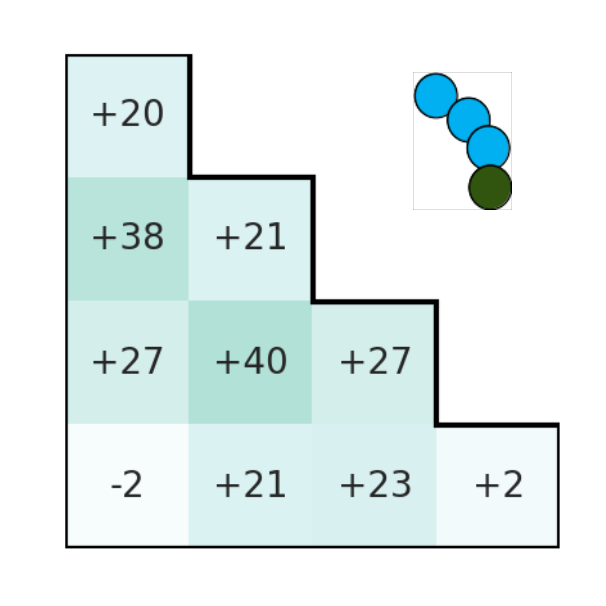}
    \caption{H1T3}
    \end{subfigure}
    ~ 
    \begin{subfigure}[b]{0.23\textwidth}
    \includegraphics[width=\textwidth]{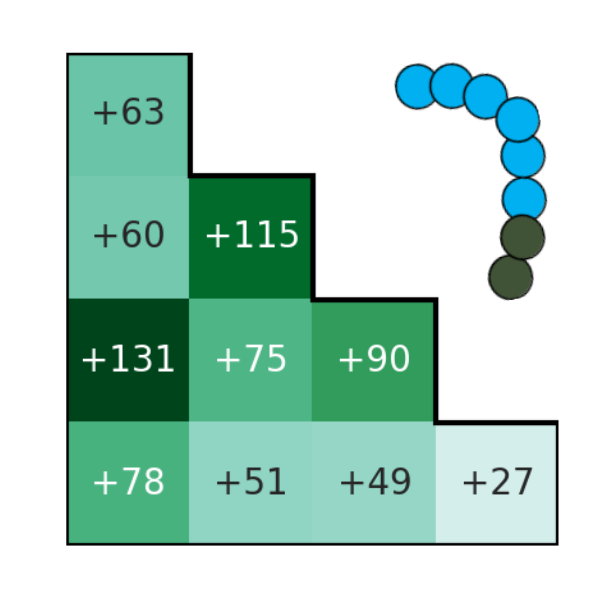}
    \caption{H2T6}
    \end{subfigure}
        \caption{Percentage increase in the simulated adsorption value above the baseline prediction, for each of the ten heterogeneous surfaces displayed in Fig.\ref{fig:surfaces}. The data is averaged across surfactant concentrations of 4.5\%, 8.5\% and 12.5\%.}
        \label{fig:triangles}
\end{figure}

The relative excess above the baseline theory is shown in Fig \ref{fig:triangles}a) and \ref{fig:triangles}b) for each of the ten heterogeneous surfaces individually displayed in Fig. \ref{fig:surfaces}.
For almost all surface patternings, adsorption slightly exceeds the baseline estimate, again more prominently for the longer molecule H6T2.  The increase is up to 6\% percent for H3T1 and up to 17\% percent for H6T2. 

The origin for this phenomenon can be uncovered by examining local adsorption densities onto the plane of the surface, as seen in an example in Fig. \ref{fig:heatmaps}-row 2 a) and b). 
While adsorption is mostly localized on the hydrophobic surface patches, it is also somewhat elevated around the boundaries of these patches. This observation suggests that adsorbed molecule chains can protrude into the space above adjoining hydrophilic beads. This steric effect can account for the minor increase in the simulated relative to the theoretical adsorbent densites, where edge effects are neglected.

Furthermore, surface (0.9, 0.9) (lower right hand corner of Fig. \ref{fig:triangles} a) \& b)) consistently stands out as one of less enhanced adsorption.  This surface has the largest hydrophilic domains, with less boundary area or hydrophilic inclusions than the other surfaces.  There is little systematic variation between the different surface patterns amongst the remaining nine heterogeneous surfaces. 

In contrast with H3T1 and H6T2, the adsorption of
H1T3 and H2T6, with their longer 
tails, is dominated by cooperative effects. The kinetics of micelle formation, dissolution, and diffusion are also slower relative to H3T1 and H6T2. As a result, surface adsorption evolves via a series of discrete events where a bulk micelle collides and sticks to the surface.  While the surface adsorbent density measured in each simulation 
passes our statistical tests for equilibration and stationarity as outlined in Subsection \ref{ssec:equilibration}, the values may still be quantitatively inaccurate if we do not compute the correct equilibrated bulk micelle size distribution.  As such, the large variation in adsorption between similar systems highlighted in Fig. \ref{fig:triangles} c) \& d), can be more readily explained by insufficient sampling as opposed to any real nonlinear effects of concentration or surface domain size.

With H1T3 and H2T6, any potential trends in adsorption with concentration are obscured by large stochastic variation between simulations. Unlike at the lowest surfactant concentration of $0.5\%$, there are large random variations in surface coverage between the simulations at $4.5\%$, $8.5\%$, and $12.5\%$ with no trend towards increased adsorption with increased concentration.  This serves as a motivation for why we have averaged over the three higher concentration simulations.

The adsorbent density for the hydrophobic molecules measured on the ten distinct heterogeneous surfaces generally exceeded the corresponding theoretical linearly extrapolated predicted values. (Fig. \ref{fig:triangles}c \& d). While the individual recorded values are expected to display large stochastic variations owing to the slow micelle dynamics inherent in such simulations, the overall percentage increases illustrated in Fig. \ref{fig:triangles} are qualitatively more significant for the hydrophobic relative to the hydrophilic molecules.

The origin of this phenomenon is the formation of surface adsorbed micelles which span the hydrophilic domains locating in the gaps between distinct hydrophobic patches. This effect is amplified for the longer H2T6 molecule, which due to steric effects, as with H6T2, protrudes its molecule chain further over adjoining hydrophilic patches. This raises the probability of two adsorbed surfactant assemblies, adsorbed on spatially separated hydrophobic domains, connecting to one another. This concept will be studied in greater detail in the next subsection.

\subsection{Localized Cooperative Adsorption}

\begin{figure}[ht]
\centering
    \begin{subfigure}[b]{\textwidth}
    \includegraphics[width=\textwidth]{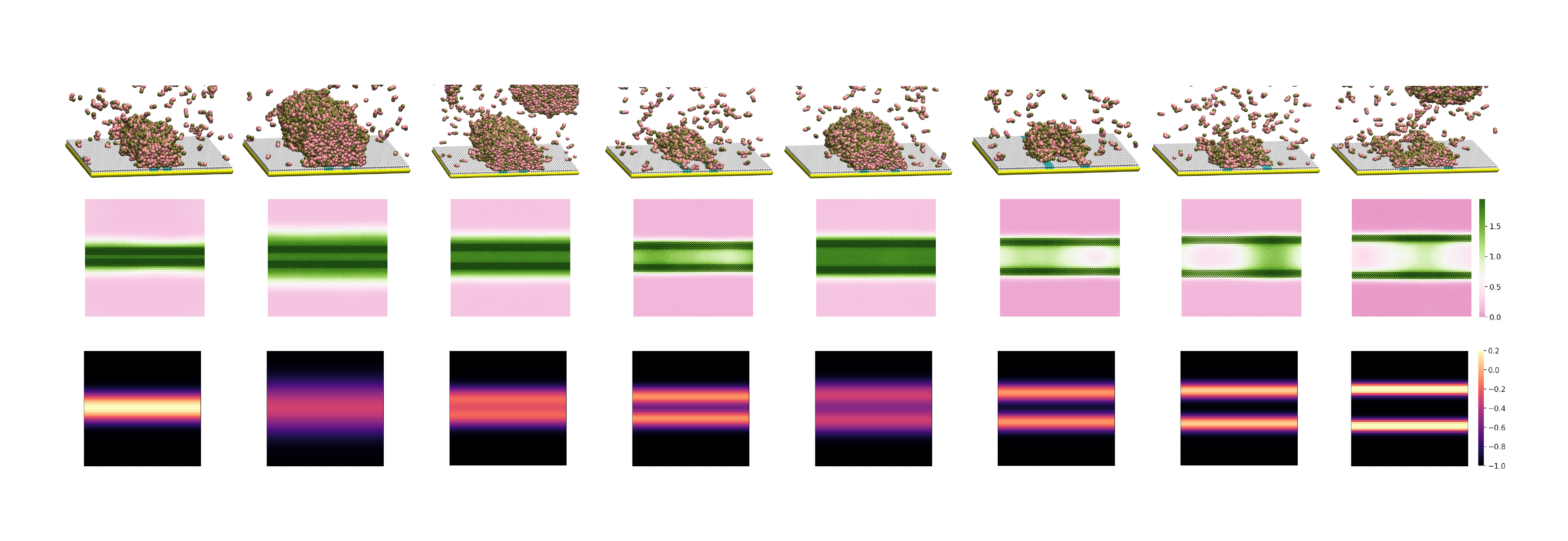}
    \caption{H1T3}
\end{subfigure}
\begin{subfigure}[b]{\textwidth}
    \includegraphics[width=\textwidth]{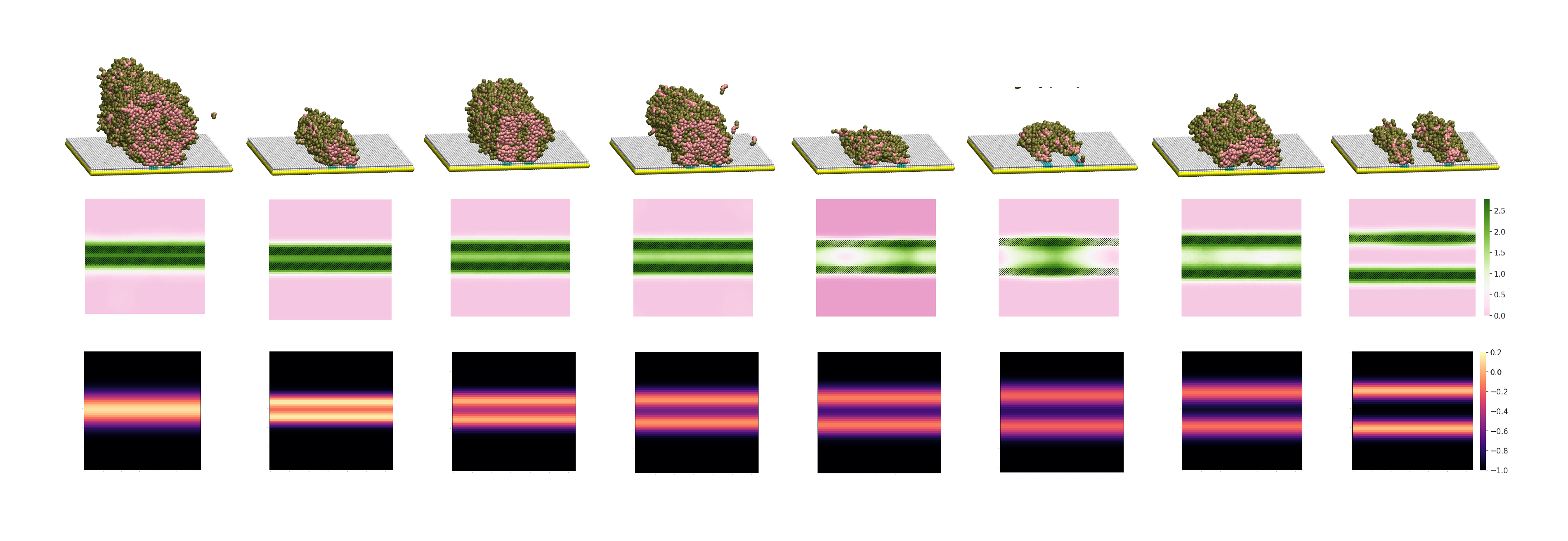}
    \caption{H2T6}
    \end{subfigure}
     \caption{Plots illustrating the cooperative adsorption of surfactants onto two parallel hydrophobic stripes.  Simulation snapshots of the adsorbed structures are shown in the top row of each subfigure. The temporally-averaged local adsorbant density distributions are displayed in the middle row, while the bottom row represents our estimated effective surface potential.  As stripe separation increases from left to right, the cooperative characteristics (connectivity and shape) of the adsorbed structures diminish.}
     \label{fig:stripes}
\end{figure}

Next, we examine how the subtle interplay between bulk micelles, bare surface, and adsorbed micelle potentials influences cooperative adsorption. 
A simple surface consisting of two attractive parallel hydrophobic stripes separated by a repulsive hydrophilic domain was chosen as the ideal template to interrogate this phenomenon.
A rich variety of adsorbent micelle structures are formed for primarily hydrophobic molecules (H3T1 and H6T2) in Fig. \ref{fig:stripes}.  The size of the individual adsorbate structures varies in a stochastic manner between individual simulations, and is related to the number of constituent molecules present in slow-moving bulk micelles when they collide with the surface. The adsorbed micelles quickly restructure into the most energetically favourable configuration following the initial collision, as determined by the heterogeneous hydrophobic-hydrophilic surface potential.

The adsorbed structures bridge the hydrophilic gap at short stripe separations as the repulsive surface potential of the hydrophilic domain is weaker than the intrinsic attractive cohesive energy of the adsorbed micelle. However, this relationship is reversed at long separations, as the repulsive hydrophilic surface potential prohibits two separate hemicylinderal adsorbed micelles from connecting with one another. The shape of the adsorbed structure also evolves as the separation increases. The most energetically favourable configuration for a micelle adsorbed on either one stripe or two adjacent stripes is a hemicylinder. However, as the repulsive hydrophilic gap grows, the competition between both forces leads to the formation of both non-continuous flat adsorbent and curved bilayer structures. 

\begin{figure}[ht]
    \centering
    \includegraphics[width=\textwidth]{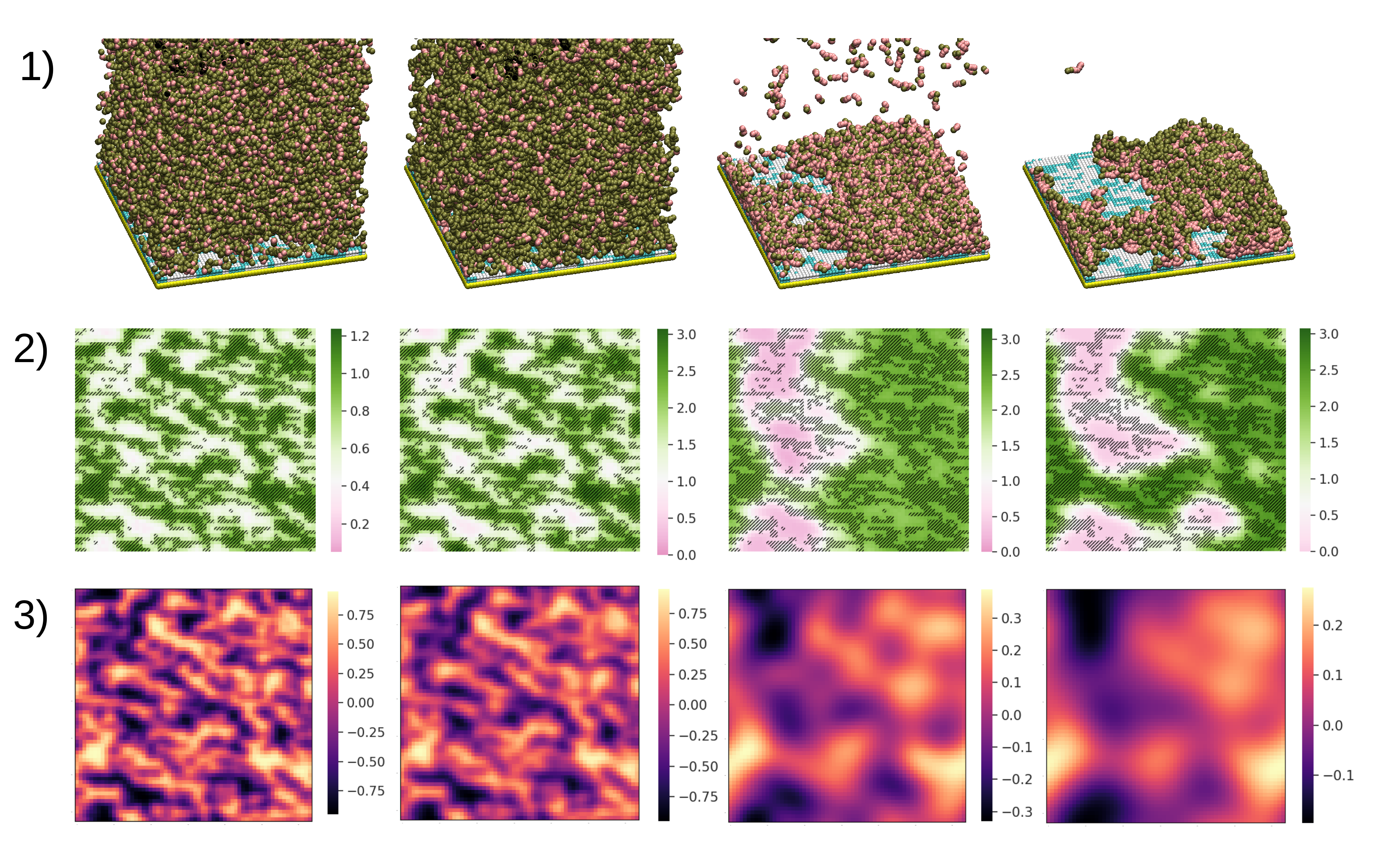}
    \caption{Plots illustrating localised variations in surfactant adsorption on the heterogeneous surface with generating probabilities (0.3, 0.6). The molecules and their respective concentrations are from left to right: H1T3 (4.5\%), H2T6 (4.5\%), H3T1 (12.5\%) and H6T2 (12.5\%). Row 1) Snapshots of the equilibrated adsorbent distributions; 2) Contour plots of the local mean adsorption density, where the pink-green colorscale signifies the  local adsorbate density relative to the global mean (white). The hatched area indicates the location of hydrophobic surface beads. 3) The effective surface potential is estimated by filtering the exact surface potential with a Gaussian filter, with the Gaussian width $\sigma$ fitted to best describe the combined measured adsorption distribution across concentrations.}
    \label{fig:heatmaps}
\end{figure}

Examining the adsorption of $\mathrm{H1T3}$ and $\mathrm{H2T6}$ onto heterogeneous surfaces, the adsorbent density is observed to be consistently high or low at certain localised regions within each surface (Figure \ref{fig:heatmaps}-2). The adsorption characteristics on hydrophobic stripes along with localized adsorption on heterogeneous surfaces indicate that there is an effective surface potential with spatial variation on a scale beyond individual surface beads. Adsorption does not follow from the `bare' potential of a grid of surface beads with interactions range limited to $r_c=1$, but rather from a more long-range effective potential which encapsulates the otherwise complex mediation of the surface potential by surfactant molecules.

As an approximation to the effective surface potential, we map the exact surface potential, as determined by the distribution of surface beads (assigned potential values $\pm1$), to a continuous smooth potential by applying a Gaussian blur.  The convolution with a Gaussian 
\begin{equation}
G(x) = \dfrac{1}{\sqrt{2 \pi \sigma^2 }} \exp\left({\frac{-\mathbf{r}^2}{2\sigma^2}}\right)
\end{equation} 
serves to filter out high-frequency variations in the surface potential. The standard deviation $\sigma$ for which the standardized smoothed potential best fits the standardized local adsorption density was numerically computed for each striped simulation (Table \ref{tab:stripesigmas}).

A comparison with the observed adsorbed micelles illustrated in Fig. \ref{fig:stripes} suggests that large values of the smoothing parameter $\sigma$ correlate with larger adsorbed micelle structures and bridging between the two hydrophobic stripes. 
\begin{table}[h!]
  \begin{center}
    \begin{tabular}{c|c|c}
      Gap Width & H1T3 & H2T6 \\
      \hline
      1.25  & 2.40 & 2.76 \\
      2.5 & 5.21 & 1.59  \\
      3.75  & 2.51  & 1.86 \\
      5 & 2.03 & 2.02 \\
      6.25 & 2.94 & 2.16\\
      7.5 & 1.99 & 2.36\\
      8.75  &1.69& 1.62\\
      10 & 1.22 & 1.76 \\
      \end{tabular}
  \end{center}
  \caption{Table detailing the Gaussian blur standard deviation $\sigma$ (in DPD length units) which best fits the adsorbed molecule density for each striped simulation.}
  \label{tab:stripesigmas}
\end{table}

Returning to adsorption onto our randomly patterned heterogeneous surfaces, we notice the emergence of distinctive intra-surface variation: certain local patches of each surface are consistently preferred or avoided for the adsorption of micelles of predominantly hydrophobic molecules, as shown here for the example of surface (0.3, 0.6) (Figure \ref{fig:heatmaps}- 2c and 2d). The spatial adsorption patterns suggest a locally varying effective surface potential. We apply a Gaussian blur to the exact surface potential, as indicated by the distribution of hydrophilic and hydrophobic surface beads. A correspondence is observed between the potential blurred to a certain level $\sigma$ and the observed local variations in adsorption (Figure \ref{fig:heatmaps}- rows 2 and 3). By numerically fitting the adsorbed surfactant density distribution to the surface-bead-derived smoothed potentials, we extract a value $\sigma$ for the adsorption patterns of each surfactant molecule (Table \ref{tab:sigmasfull}). 

\begin{table}[h!]
  \begin{center}
    \begin{tabular}{c|c|c|c|c}
      Surface ($p_x, p_y$)& H3T1&H6T2& H1T3 & H2T6 \\
      \hline
      (0, 0)  & 0.76 & 0.84 &4.77 & 4.78 \\
      (0, 0.3) & 0.90 & 0.89 &3.86 & 4.17 \\
      (0, 0.6)  & 0.73  & 0.88 & 3.03 & 3.38\\
      (0, 0.9)& 0.71 & 0.86 & 2.21 & 2.61 \\
       (0.3, 0.3) & 0.73 & 0.89& 3.87& 3.29\\
       (0.3, 0.6) & 0.73 & 0.89 & 3.93 & 2.72\\
       (0.3, 0.9)  &0.74 & 0.89 & 3.23 & 2.24\\
       (0.6, 0.6) & 0.76 & 0.93 &2.26 & 2.04\\
       (0.6, 0.9)& 0.75  & 0.93 & 1.58& 1.89\\
       (0.9, 0.9) & 0.74 & 0.92 &2.14 &  1.68\\
      \end{tabular}
  \end{center}
  \caption{Table listing the Gaussian blur standard deviation $\sigma$ that best fits the measured adsorption pattern on each heterogeneous surface. Note that the adsorption distributions were first averaged for all equilibrated frames and then across each of the four simulated bulk molecule concentrations.}
  \label{tab:sigmasfull}
\end{table}

For the molecules observed to form micelles at the concentrations and time-scales explored in this work, namely $\mathrm{H3T1}$ and $\mathrm{H6T2}$, large stochastic variations between individual simulations with the same surface but at different molecule concentrations obscure any clear concentration-dependent trend. Instead, a superior analysis of the effective surface potential can be achieved by averaging local adsorption across the series of four simulations at different concentrations. Considering all observed localised adsorption distributions cumulatively, including at the lowest concentration, has the desirable effect of giving extra weight to the most attractive domains that consistently adsorb molecules at low concentration.

The gradients of the effective surface potential experienced by the bulk surfactants vary significantly as a function of their hydrophilic:hydrophobic ratio, as highlighted in Table \ref{tab:sigmasfull}. Predominantly hydrophilic surfactants ($\mathrm{H3T1}$ \& $\mathrm{H6T2}$) recorded small values for the Gaussian blur width (less than unity for all surfaces). This indicates that adjacent molecules feel a short-ranged effective surface potential akin to the cumulative interactions of the surface bead types. This similarity indicates that independent adsorption mechanisms dominate and cooperative effects are negligible.

Conversely, the mainly hydrophobic surfactants ($\mathrm{H1T3}$ \& $\mathrm{H2T6}$) are subjected to gently varying effective surfactant potentials, as evidenced by the wide Gaussian blur widths reported in Table \ref{tab:sigmasfull}. Long-ranged adsorbed micelle cooperative effects govern the many-body effective surface potential which deviates significantly from the cumulative interactions of the surface bead types and facilitates adsorbent bridging between hydrophobic domains.

One of main results of this paper is that the longer the molecule backbone length, the greater the probability for adsorbed micelles to bridge between hydrophobic regions, as previously discussed in subsection \ref{ssec:hydrophobic}. In the first two columns of Table \ref{tab:sigmasfull}, this observation translates into a greater $\sigma$ value for the longer $\mathrm{H6T2}$ molecule relative to the shorter $\mathrm{H3T1}$ molecule.
In theory, this effect should also be evident within the hydrophobic columns, however the stochastic noise emanating from the slow dynamics and micelle formation inherent in these simulations obscures this phenomena. Hence, the calculated $\sigma$ values for $\mathrm{H2T6}$ do not always exceed the corresponding readings for $\mathrm{H1T3}$.

A visible gradient in the values of $\sigma$ is observed across the heterogeneous surfaces, originating from the spatial extent of the patches documented in Table \ref{tab:numMeanLength}. This trend is detected for the predominantly hydrophobic surfactants, where the $\sigma$ values decrease as one increases the spatial order in either or both dimensions. This follows from the corresponding decline in surface patch boundaries, where the most significant transitions in the effective surface potential are expected to be located.

For molecules with strong cooperative interactions, adsorption onto the surface is laterally mediated by other molecules.  Therefore, the effect of surface detail below a certain threshold length scale is diminished when molecules form cooperative surface structures.  Instead, the success of our Gaussian blur analysis lies in the realisation that it is the low-frequency variations in the surface structure itself that determines the local variations in adsorbent density distributions.

\section{Surface Patterning for Calibrating Adsorption} 

The methods we have described can be generalised to help scientists to tailor heterogeneous two-component surfaces, with fixed global composition ratio, to either inhibit or amplify molecular adsorption for industrial applications.
In this scenario, the adsorption coverage is determined by an effective surface potential which a bulk molecule experiences in the vicinity of the solid-liquid interface. The 2D surface morphology which maximises global adsorption must derive the optimal balance between the depth and breadth of adsorption wells in this potential landscape. If the attractive domains are too small, then the many resulting potential wells will be narrow and shallow, resulting in the rapid detachment of the adsorbed material. Similarly, if the domains are too large, adsorption will be confined to the few large attractive domains and the global coverage will be bound by this limit.
\par
The surface topologies where the overall adsorption is maximised can be located in a `Goldilocks' region in between these two limits. The precise position of this region in parameter space will vary from system to system and will depend upon the complex interplay between the adsorbant molecules, their interactions with the two surface materials and their self-assembly dynamics. 
\par
In order to locate the optimal surface topology, a large number of different surfaces need be generated and tested.
This problem can be transformed into a two-dimensional surface parameter optimisation exercise where one fits the generating probabilities $p_{x}$ and $p_{y}$, while the target surface area fraction $s_{tar}$ remains invariant.
This step opens up the possibility of applying automated statistical algorithms to efficiently drive this {\it{in silico}} exploration of accessible surface configurations\cite{hernandezlobato2017}. 

A similar approach may also be adopted when the objective is to achieve a bespoke adsorption pattern on the surface. In this case instead of maximising or minimising the total number of adsorbed molecules, we optimize the surface topology to reproduce the desired localized adsorption characteristics. Specifically, we can define an objective function which compares the simulated spatial adsorbant density distribution to the desired pattern and then explores the space in the same way as outlined above. This is also complementary to existing experimental methods investigating the inverse design problem, which qualitatively infer the surface topology configuration from the measured adsorption characteristics \cite{Zhang2007}.

\section{Conclusions}

Our work extends the types of surfaces that can that can be studied via simulation beyond regularly patterned surfaces, likes stripes checkerboards, to more realistic randomly disordered surfaces.
This is facilitated by our  surface generation algorithm that stochastically generates flat dual-component surface typologies as a function of three distinct spatial disorder parameters (hydrophobic/hydrophilic surface area ratio and neighbour similarity along the x and y axis). 
\par
We used our algorithm to perform a theoretical examination of the competing phenomena that determine the adsorption of a set of model nonionic surfactants of varying lengths, hydrophobicities and concentrations.
We found the adsorption isotherms when surfactant concentration was below the CMC to be langmuir-like, matching theoretical expectations.
We observed a rich variety of adsorbed micelle morphologies, with long-ranged cooperative forces leading to the formation of assemblies spanning repulsive hydrophilic regions located in the gaps between discrete attractive hydrophobic domains.  
In general all the trends we observed can be understood as resulting from the interplay of micelle-cohesion, surface-surfacant and surfacant-surfactant interactions and the chemical topology of the surface.  
\par
We found that applying a gaussian-blur analysis to the bare surface potential of the nanopattern configuration allowed us to map it, using a one-parameter fit, to the spatial distribution of the adsorbant density. 
For the molecules and surfaces we studied we observed monotonic behaviour of this fit parameter with respect to the topological parameters we varied. 
In combination with a method for interpolating the value of the fit parameter as a function of the surface-topology parameters, this could potentially allow one to predict the adsorption distribution of the absorbant material for any given surface topology. 

\section{Acknowledgements}

This work was supported  by the STFC Hartree Centre’s \textit{Innovation: Return on Research Programme}, funded by the UK Department for Business, Energy \& Industrial Strategy. 
JK and SF acknowledge, with appreciation, support from the IBM internship scheme.

\bibliography{Paper}
\end{document}